# The route to high temperature superconductivity in transition metal oxides

A. Bussmann-Holder* and K. A. Müller**

*Max-Planck-Institut für Festkörperforschung, Heisenbergstr. 1, D-70569 Stuttgart, Germany

**Physik Institut der Universität Zürich, Winterthurerstr. 190, 8057-Zürich, Switzerland

The discovery of high temperature superconductivity in cuprates was possible only through an intimate knowledge of perovskite oxides which have been synthesized and characterized for decades at the IBM in the Zürich laboratoty. Especially $SrTiO_3$ and $LaAlO_3$ have been in the focus at IBM as was presented in Volume 1 of a series [1]. Probably for the first time detailed microscopic investigations of the local properties of these compounds have been obtained by studying by means of EPR the surroundings of transition metal impurities in these materials. These experiments enabled the identification of the order parameter of the structural instability observed in these oxides. However, the idea to search for superconductivity came later motivated by theoretical considerations that metallic hydrogen could become superconducting at high temperatures. Since $SrTiO_3$ is an insulator it was thought that the implantation of hydrogen would render it metallic and eventually also superconducting. This approach failed since the carrier density remained always too small. In sequence it was then tried to achieve a metallic state in oxide perovskites by varying their composition which was in so far promising as reduced $SrTiO_3$ exhibits superconductivity at 0.3K [2]. Furthermore it was subsequently shown [3] that $T_c$ can be enhanced to 1.2K by doping $SrTiO_3$ with Nb. In spite of the fact that $T_c$ was far below values achieved in A15 compounds, a remarkable observation was connected with the Nb doped perovskite, namely for the first time long before predicted two-gap superconductivity was realized here.

In spite of the rather disappointing low transition temperatures achieved in $SrTiO_3$ another observation, namely the discovery of $T_c$ enhancements in granular Al as compared to crystalline

Al, kept the interest in this field alive. Opposite to crystalline Al, the small metallic grains in the non-crystalline Al, are surrounded by amorphous $Al_2O_3$. These small grains couple by Josephson junctions and enhance $T_c$ by a factor of up to three, however, still remaining on the side of low $T_c$ materials [4, 5].

Early predictions by Matthias do search for high $T_c$ superconductivity in simple metallic or at most binary compounds have been followed for many years but unfortunately could not exceed values of $T_c$>23K in $Nb_3Sn$ in the early 70'th. This fact encouraged to search for other than intermetallic compound as new superconductors with special focus on oxides, actually definitely excluded from the Matthias considerations. Since in these days it was undoubtedly accepted that the BCS theory is best suited to explain superconductivity, a simple inspection of the BCS $T_c$ defining equation shows the material limits: $k_B T_c = 1.13\hbar\omega_D \exp(-1/\{N(E_F)V_{ph}\}$. In oxides the density of states at the Fermi level $N(E_F)$ is low, $N(E_F) = 4\,10^{21}/cm^3$, on the other hand the electron-phonon interaction $V_{ph}$ is rather strong. Thus it could be an interesting route to enhance $T_c$ by increasing the carrier density via appropriate doping, or by strengthening the electron-phonon interaction. Especially, it was suggested that mixed valency compounds could support an increasing $N(E_F)$, whereas polaron formation was shown to enhance $V_{ph}$. Chakraverty [6] was one of the first to calculate the phase diagram of a polaronic superconductor which has many features in common with the one of cuprate superconductors. By defining $\lambda = N(E_F)V_{ph}$, the phase diagram (Figure 1) shows metallic properties for small values of $\lambda$, whereas it is insulating bipolaronic for $\lambda$ large. In the intermediate $\lambda$ regime superconductivity is realized which may exhibit exceedingly large values of $T_c$. This quite intriguing ideas left the question open, in which systems such a scenario could be realized.

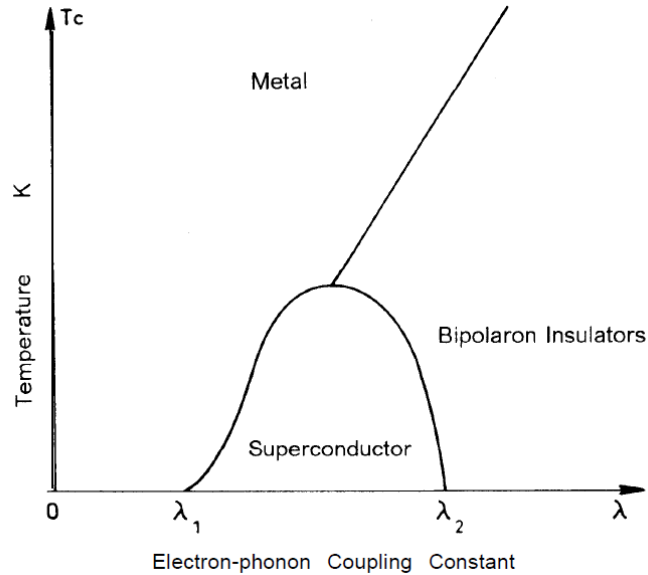

**Figure 1** Phase diagram as a function of electron-phonon coupling $\lambda$.

At this stage a new concept came into play, namely the idea of the Jahn-Teller polaron, as suggested by Höck et al [7] and discussed in more detail below. The basic ingredient of the Jahn-Teller effect is based on the instability of orbitally degenerate electronic states towards a lattice distortion whereby the degeneracy is lifted. Here a competition between electron localization and their kinetic energy sets which is dominated by the Jahn-Teller energy $E_{JT}$ versus the band width $W$. For $E_{JT} >> W$ localization takes place and the electron is trapped in a phonon cloud. In the opposite case $E_{JT} << W$ the electron travels almost unaffectedly through the lattice experiencing only small disturbances from it. The interesting case $E_{JT} \approx W$ combines the limiting cases since here the electron travels through the lattice with its own displacement field. This is the Jahn-Teller polaron. Regarding oxides, the knowledge of these established that many of them contain transition metal ions with partially filled $e_g$ orbitals which are known to act as Jahn-Teller centers. Of special interest are in this respect $Ni^{3+}$, $Fe^{4+}$ or $Cu^{2+}$ which were then considered to be possible candidates for high temperature superconductivity.

The search started in 1983 with the La-Ni-O systems. $LaNiO_3$ is metallic and the Jahn-Teller energy is smaller than the band width. In order to reduce the band width, Al was substituted for Ni where for small concentrations an increase in the resistivity sets in and a semiconductor type behavior is obtained for large concentrations turning into a localization regime with decreasing temperature. Superconductivity could not be realized by this approach. Another option to reduce the band width is strain which can be achieved via replacement of the $La^{3+}$ ion by the smaller $Y^{3+}$ ion. However, the results were comparable to the previous replacement and remained unsuccessful with respect to superconductivity.

Two years later improvement in the experimental situation enabled the final break through. Since the experiments with $Ni^{3+}$ did not reveal the expected results, $Cu^{2+}$ was considered as another candidate for the Jahn-Teller polaron effect. By partially replacing the Jahn-Teller ion $Ni^{3+}$ by the non Jahn-Teller ion $Cu^{3+}$ an increase in the sample resistivity was achieved, however, preserving its metallic character down to 4K. Shortly afterwards a report on the La-Ba-Cu-O system appeared which showed that this compound is metallic between 300 und 100 C [8]. The special property of this system is the mixed valency of Cu which is present as $Cu^{2+}$ and $Cu^{3+}$. Accordingly, it is possible to tune the Cu valency continuously by changing the La/Ba ratio. When cooling these samples a metallic like resistivity decrease was obtained followed by an increase at lower temperature which is a signature of localization. This behavior continued to 30K when a sudden drop in resistivity set in around 11K. These data could be reproduced on several samples confirming the same temperature behavior (Figure 2).

By varying the La/Ba ratio and the thermal treatment, the drop could be shifted to temperatures as high as 35K, much higher than the one observed in $Nb_3Ge$ at 25K. The origin of the resistivity drop remained open since the Meissner-Ochsenfeld effect had not been demonstrated. Nevertheless, the first paper with the carefully selected title "Possible high $T_c$ superconductivity in the La-Ba-Cu-O system" was submitted for publication [9]. In addition, it turned out that the such synthesized samples contained two different phases and it had to be verified which of them corresponded to the superconducting one. By systematically changing the composition and measuring electrical and lattice properties, the tentative assignment was made that the Ba containing $La_2CuO_4$ was responsible for superconductivity. Interestingly, a correlation between Ba doping, structural distortion and superconductivity was observed, namely, that the orthorhombic distortion decreases with increasing Ba content to make the structure more tetragonal with the maximum $T_c$ almost coinciding with the tetragonal to orthorhombic transition.

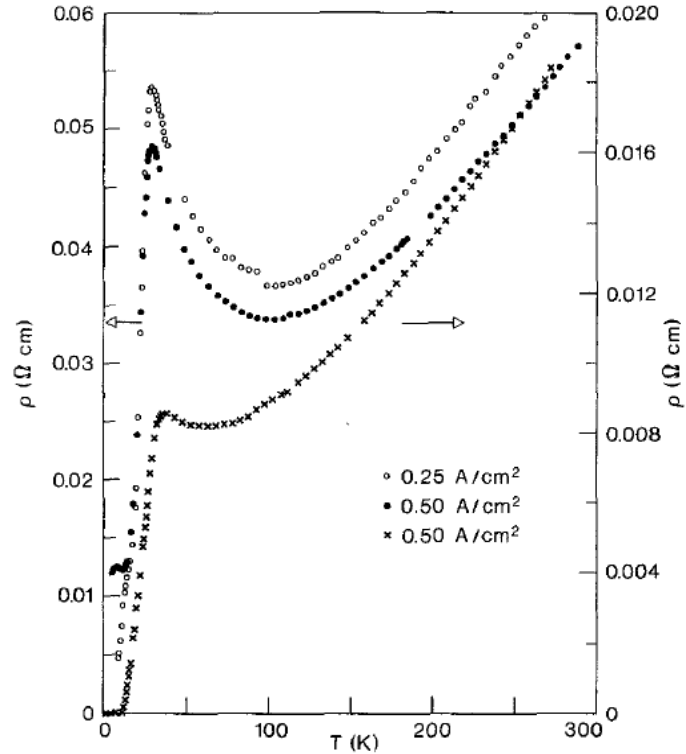

**Figure 2** Temperature dependence of the resistivity in $Ba_xLa_{5-x}Cu_5O_{5(3-y)}$ for samples with x(Ba)=1 (upper curves, left scale) and x(Ba)=0.75 (lower curve, right scale). The first two cases also show the influence of current density (after Ref. 9).

In September 1986 it was possible to carry out magnetic measurements where first a compound with low Ba content was measured. In this system metallic conductivity was observed down to 100K followed by a transition to localization. The susceptibility was Pauli like positive, temperature independent, changing to Curie-Weiss behavior at low temperatures. Samples exhibiting a resistivity drop underwent a transition from paramagnetic to diamagnetic, characteristic of superconductivity-related shielding currents. The diamagnetic transition always started slightly below the resistivity drop, evidencing percolative superconductivity. The final proof for superconductivity, namely the Meissner-Ochsenfeld effect, had thus been demonstrated and the superconducting phase identified (Figure 3).

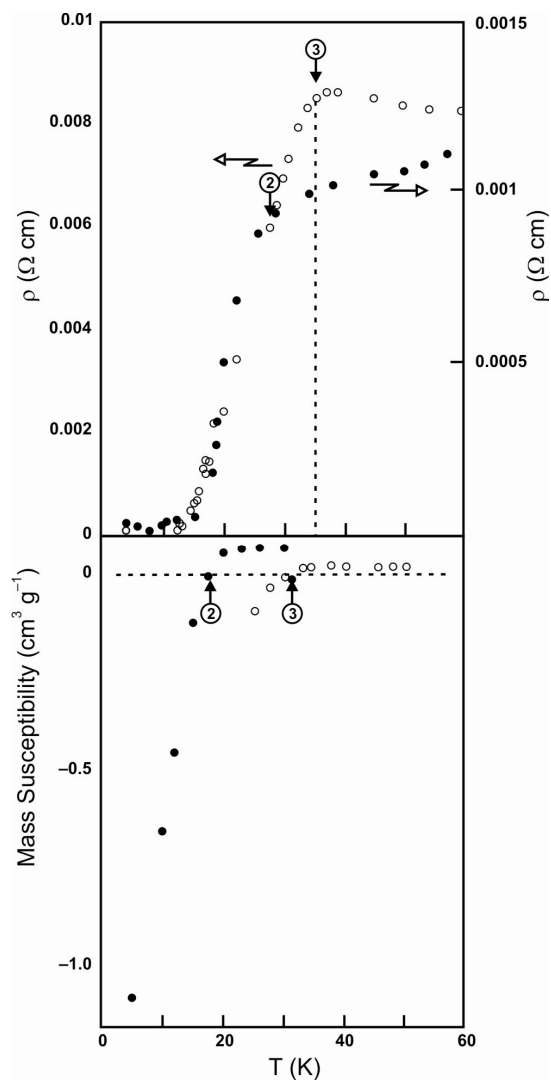

**Figure 3** Low-temperature resistivity and susceptibility of (La-Ba)-Cu-O samples 2(●) and 3 (○)

from Ref. 10 and 11. Arrows indicate the onset of the resistivity and the paramagnetic to diamagnetic transition, respectively.

In the following, the magnetic characterization of the samples was continued and evidence for a glass state discovered [12]. More important was, however, the idea to replace La not only by Ba but to try substitutions with Sr and Ca. Sr substitutions induced superconductivity at even higher temperatures than Ba with a maximum onset of 40K [13]. It is worth mentioning that the radius of Sr is almost identical to the one of the La ion and consequently this ion fits better into the structure than Ba.

Even though the above results created an enormous positive international response, the breakthrough was the confirmation of the data by other groups. At the end of the year 1986 a Japanese group reproduced the data [14] and the community which had been skeptical before, became very attentive.